\documentstyle[emulateapj]{article}
\begin{document}
\title{A Shotgun Model for Gamma Ray Bursts} \author{Sebastian
  Heinz\altaffilmark{1} \and Mitchell C.\ Begelman\altaffilmark{2}}
\affil{JILA, University of Colorado and National Institute of Standards
  and Technology, Boulder, Colorado 80309-0440\altaffilmark{3}}
\altaffiltext{1}{email address: heinzs@rocinante.Colorado.edu}
\altaffiltext{2}{email address: mitch@jila.Colorado.edu}
\altaffiltext{3}{also at Department of Astrophysical and Planetary
  Sciences, University of Colorado, Boulder}
\begin{abstract}
  We propose that gamma ray bursts (GRBs) are produced by a shower of heavy
  blobs running into circumstellar material at highly relativistic speeds.
  The gamma ray emission is produced in the shocks these bullets drive into
  the surrounding medium.  The short term variability seen in GRBs is set
  by the slowing-down time of the bullets while the overall duration of the
  burst is set by the lifetime of the central engine.  A requirement of
  this model is that the ambient medium be dense, consistent with a strong
  stellar wind.  The efficiency of the burst can be relatively high.
\end{abstract}
\section{Introduction}
Gamma ray bursts (GRBs) show variability over a large range in time scales
--- from millisecond spikes in BATSE light curves (Bhat et al.\ 1992) to
months in the associated afterglows (Bloom et al.\ 1999, Frail et al.\ 1997).
The discovery of afterglow redshifts that place GRBs at cosmological
distances implies isotropic energies of $E_{\rm iso} \gtrsim 10^{53}\,{\rm
  ergs}$ for the average burst in gamma rays alone, released over an
observed time of order $T_{\rm obs} \sim 10\,{\rm s}$.  Since GRB spectra
are believed to be optically thin, the only viable explanation is a highly
relativistic outflow that releases its energy at sufficient distances from
the central engine for the optical depth and the compactness of the plasma
to be small, with required bulk Lorentz factors of $\Gamma \gtrsim 100$.

In early models (M$\acute{\rm e}$sz$\acute{\rm a}$ros \& Rees 1992, Katz
1994a, Sari \& Piran 1995) the gamma rays were produced by the shock such
an outflow drives into the interstellar medium (thus called `external shock
models').  While providing a way to produce the required energy in gamma
rays over the observed GRB durations, these models have not been very
successful at explaining the observed short term variability: in order to
see strong variability, the surrounding medium must be very clumpy (e.g.,
Shaviv \& Dar 1995, Fenimore et al.\ 1996). It has been argued that this
process might lead to very low efficiencies at converting kinetic into
thermal energy (Sari \& Piran 1997). However, a quantitative investigation
casts some doubt on this argument (Dermer \& Mitman 1999).

These difficulties of the external shock model prompted Sari \& Piran
(1997) to postulate that the gamma ray emission must instead be produced by
the internal shock scenario (Narayan, Paczy$\acute{\rm n}$ski, \& Piran
1992, Rees \& M$\acute{\rm e}$sz$\acute{\rm a}$ros 1994).  In this picture,
the duration of the burst $T$ is set by the time scale over which the
central engine operates, while the substructure in the bursts on time
scales $\tau$ is produced by inhomogeneities in the outflow.  These
inhomogeneities are assumed to travel at different bulk Lorentz factors
$\Gamma$.  Upon running into each other, these shells of material get
shocked and release some of their kinetic energy in the form of gamma rays.
This picture has become the paradigm in GRB physics.  However, recent
estimates of the energy conversion efficiency $\eta$ indicate that at best
a few percent of the bulk kinetic energy carried by the outflow can be
converted into gamma rays in internal shocks, which leads to uncomfortably
high requirements on GRB energies (Panaitescu, Spada, \& M$\acute{\rm
  e}$sz$\acute{\rm a}$ros 1999, Kumar 1999; larger efficiencies can be
achieved if you assume large dispersion in $\Gamma$, Katz 1997). A
non-spherical geometry can reduce the required energy, however, a very
small opening angle of the outflow implies a high rate of unobserved GRBs,
which is hard to reconcile with the number of observed supernovae (which
are believed to produce GRB precursors - either compact objects or
hypernovae). See Piran (1999) for a detailed review of the subject.

Fenimore et al.\ (1996, 1999, see also Woods \& Loeb 1995) recently
suggested that an external shock scenario could give rise to the observed
variability if the spherical symmetry of the outflow were broken, still in
the context of what Sari \& Piran (1997) call a `Type I' model, i.e., the
burst duration is set by the slowing-down time of the ejecta.  However, the
observed temporal constancy of the pulse width in individual spikes of
GRB990123 seems to rule out such a model (Fenimore, Ramirez-Ruiz, \& Wu
1999. This would also argue against the scenario suggested by Dermer \&
Mitman, 1999.) We propose a different way by which substructure in the
outflow can produce a GRB, also via the interaction with the external
medium, but in a `Type II' scenario, i.e., the duration of the burst is set
by the lifetime of the central engine (see also Chiang \& Dermer 1999,
Blackman, Yi, \& Field 1996).

In this model, the outflow itself is very clumpy, with most of the energy
concentrated in small blobs, which are sprayed out with high $\Gamma$ over
a small opening angle.  These bullets then slam into the surrounding
medium (not unlike a meteor shower or a shotgun blast), where they release
their kinetic energy and produce gamma ray emission via external shocks,
as described in \S2.  Note that, in this paper, we assume the existence of
bullets; their possible origin is a topic for further research (see \S3
for comments). Section 3 contains a discussion on radiative efficiencies,
simulated burst light curves, and afterglows, and \S4 contains a brief
summary. 

\section{The Gamma Ray Burst Model}

BATSE light curves of most GRBs exhibit very spiky substructure (Fig.\ 1).
In the context of our model, these spikes are produced by individual
bullets of cold ejecta slamming into the surrounding medium.  As we will
show, a distribution of masses and/or Lorentz factors of these bullets can
reproduce the observed signatures of GRBs reasonably well.
  
{\plotfiddle{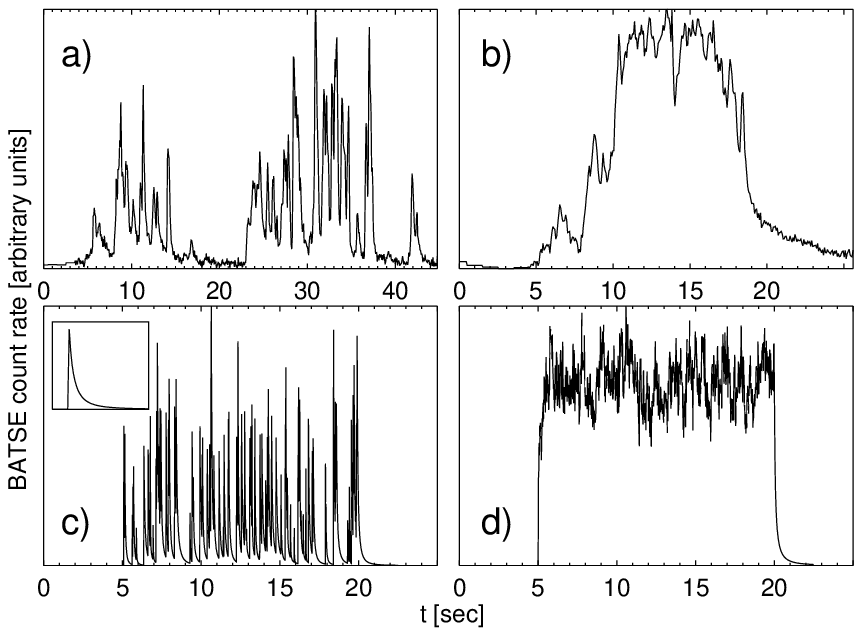}{2.5in}{0}{100}{100}{-125}{0} 
\figcaption{BATSE light curves for GRB920627 (panel a) and GRB980329 (panel
  b) and two synthetic light curves.  These curves were calculated for a
  burst duration of $15\,{\rm s}$, a mass distribution of $N(M_{\rm b})
  \propto {M_{\rm b}}^{-1/3}$, an average slowing down time of $\tau =
  0.01\,{\rm
    s}$, and $N_{100}=1$ (panel c), $N_{100}=100$ (panel d).  The insert
  in panel c shows a template light curve for a single shot.
  \label{fig:lightcurves}}}
\bigskip

In the following we assume that the central engine of the burst releases a
number of bullets $N'$ distributed over a fan of opening angle of $\theta
\sim 10^{\circ}$ with Lorentz factor $\Gamma \equiv 1000\,\Gamma_{3}$ and
released over a time period of $T \sim 10-100\,{\rm s}$.  Each bullet is
assumed to freely expand sideways, with a sideways velocity of $v_{\perp} =
\alpha/\Gamma \equiv 10^{-2}\alpha_{-2}/\Gamma \ll 1/\Gamma$, measured in
the observer's frame ($\alpha$ is the sideways velocity in the comoving
frame).  The assumption that $\alpha \ll 1$ implies that the internal
expansion speed is very sub-relativistic.

Since $\theta \gg 1/\Gamma$, we only see a fraction of the total released
energy, $E_{\rm obs} = 10^{47} \cdot E_{53}/{\Gamma_{3}}^2\,{\rm ergs}$
(where $E_{53}$ is the inferred isotropic energy in units of $10^{53}\,{\rm
  ergs}$) and an observed number of bullets $N \equiv 100\cdot N_{100} =
N'/(\theta^2 \Gamma^2)$.  It is essential in our model that the covering
fraction be less than unity (otherwise it would turn into an internal shock
model), thus we require $\alpha < 1/\sqrt{N}$.

A bullet of mass $M_{\rm b} = 5.5\times 10^{-13} E_{53} M_{\odot} / N_{100}
{\Gamma_{3}}^3$ will have converted half of its kinetic energy into
internal energy (which can subsequently be radiated away as gamma rays, see
below) after it has swept up or ploughed through a column of interstellar
gas of mass $M_{\rm s} \sim M_{\rm b}/\Gamma$.  If the duration of the
observed spike is $\tau_{\rm s} \equiv 10^{-3} \tau_{-3}\,{\rm s}$, then
the length over which the material is swept up must be of the order of
$\Delta R = 3 \times 10^{13} \tau_{-3} {\Gamma_{3}}^2\,{\rm cm}$.  If the
ambient density is $n_{\rm amb} \equiv 10^{8} n_{8}\,{\rm cm^{-3}}$, the
required Lorentz factor is
\begin{equation}
  \Gamma = 2300\cdot\left({\frac{E_{53}}{{\alpha_{-2}}^2 N_{100}
  {\tau_{-3}}^{3} n_{8}}}\right)^{1/8}
\end{equation}

This model fails for low ambient densities, as has already been discussed
in the literature (e.g., Sari \& Piran 1997).  However, if the surrounding
medium is very dense, $n \sim 10^{8}\,{\rm cm^{-3}}$, Lorentz factors of
$\Gamma \sim 1000$ can explain the observed short term variability.  The
immediate conclusion is that in this model GRBs are {\em not} caused by
mergers of naked compact objects.  Rather, the required high ambient
densities tie this model to the hypernova picture (Paczy$\acute{\rm n}$ski
1998, Woosley 1993), which {\em predicts} that the material surrounding
the blast is dense because of the pre-hypernova stellar wind. (Dense
circum-GRB matter was also suggested in a different context by Katz,
1994b.)

The fate of the outer layers of a hypernova precursor is unknown.  If the
bullets have to travel through a significant fraction of the star's mantle
(which is optically thick and thus {\em not} the site where the gamma rays
are produced), their opening angle must be extremely narrow: $\alpha <
5\times 10^{-4} E_{53}/(\sqrt{N_{100}}\Gamma_{3} M_{\rm M})$, where $M_{\rm
  M}$ is the mass of the mantle in units of $M_{\odot}$.  Since hypernovae
are believed to originate from rapidly rotating massive stars collapsing
into a compact object, it is possible that the material along the rotation
axis has collapsed before the GRB, in which case the bullets would travel
freely until they hit the circumstellar material.

Similarly, little is known about the conditions of the pre-hypernova
circumstellar material other than that it must be dense.  Massive stars are
known to have strong winds with mass loss rates of $\dot{M} \sim
10^{-6}-10^{-4} M_{\odot}{\rm yr^{-1}}$ and wind velocities from $v_{\rm
  w}\sim 20\,{\rm km\,s^{-1}}$ (red supergiant) to $v_{\rm w} \sim
1000\,{\rm km\,s^{-1}}$ (blue supergiant).  These winds must still be
present after the star collapses.  In the following, we will assume that
the GRB is produced in this leftover wind.  The density profile in the
ambient matter, then, roughly goes as $n \propto r^{-2}$ outside some
radius $R_0$.  It is natural to assume that $R_0$ is of the order of the
stellar radius, $R_0 \sim 10^{12}\,{\rm cm}$ for a blue supergiant and $R_0
\sim 10^{14} \,{\rm cm}$ for a red supergiant.  As a conservative estimate
we assume that the sphere inside $R_0$ is evacuated.

If we parameterize the density as 
\begin{equation}
  n_{\rm amb}(r > R_0)=\frac{1.5\times
  10^{36}\,{\rm cm^{-3}}}{r^2}\frac{{\dot{M}}_{-6}}{v_{20}}
\end{equation}
where $\dot{M}_{-6}$ is the mass loss rate in units of $10^{-6}
M_{\odot}\,{\rm yr^{-1}}$, $r$ is in cm, and $v_{20}$ is the wind velocity
in units of $20\,{\rm km\,s^{-1}}$, the observed slowing-down time scale is
given by
\begin{equation}
  \tau = 0.05\,{\rm
  s}\frac{E_{53}v_{20}}{{\alpha_{-2}}^{2}N_{100} {\Gamma_{3}}^{4}
  \dot{M}_{-6}},
  \label{eq:timescale}
\end{equation}
independent of $R_0$.  Thus $\tau$ can be of the order of a few
milliseconds for both red supergiant and blue supergiant winds if $\Gamma
\sim 1000$.  However, the observed time scale could conceivably be longer
than this.  The angular smearing time scale $\tau_{\rm ang}$ is defined as
the spread in light travel time to the observer across the emitting
surface.  For a bullet at a viewing angle of $1/\Gamma$ this gives
$\tau_{\rm ang} \sim \alpha (R_0 + \Delta R)/\Gamma^2$, which is longer
than $\tau$ if $R_0 \gtrsim \Delta R/\alpha = 1.5 \times 10^{17}\,{\rm
  cm}\cdot E_{53} v_{20}/({\Gamma_{3}}^2 {\alpha_{-2}}^{3} N_{100}
\dot{M}_{-6})$.  This is only of concern for {\em very} dense red
supergiant winds, and only if the region interior to $R_0$ is evacuated.

In Fig.\ 2 we show various limits on $n_{\rm amb}$ and $\Gamma$ for a fixed
opening angle of $\alpha_{-2} = 1$:
\begin{itemize}
\item[a)]{Each bullet is expected to plough through undisturbed medium.
    Thus, the covering fraction $N {\alpha}^2$ of all the bullets together
    must be smaller than 1.  For a slowing down time of $\tau_{-3} = 1$,
    this gives the dashed line in the plot, to the left of which the
    covering fraction is larger than unity.}
\item[b)]{The material between the location where the bullets release their
    energy and the observer must be optically thin.  For $\tau_{-3} = 1$
    and for $R_0 = 0$ (the most conservative limits) this constraint
    produces the dash-dotted line to the left of which the optical depth is
    larger than unity.}
\item[c)]{The forward shock must be radiative (see \S\ref{sec:discussion}).
    This constraint is shown as a light grey region inside of
    which the shock is not radiative.}
\item[d)]{The angular smearing time $\tau_{\rm ang}$ must be smaller than
    the observed stopping time $\tau$.  This limit is shown as a dotted
    line for $R_0 = 10^{14}\,{\rm cm}$.  To the right of this line, the
    smearing time is longer than the observed slowing down time.}
\end{itemize}  
The hatched region in the plot shows how the allowed region of parameter
space opens up if we relax the time scale requirements to $\tau_{-3} =
10$. 

{\plotfiddle{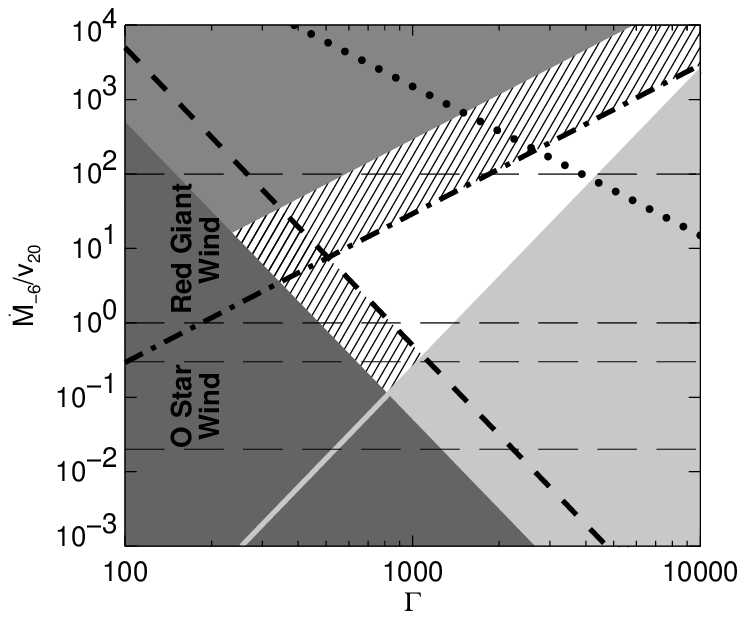}{3in}{0}{115}{115}{-131}{0}
\figcaption{Constraints on the ambient density and $\Gamma$ for $E_{53}=1$,
  $N_{100}=1$, and $\alpha = 0.01$. a) covering factor larger than unity:
  left of dashed line for $\tau_{-3}=1$, dark grey area for $\tau_{-3}=10$.
  b) optical depth between bullet and observer larger than unity: left of
  dash-dotted line for $\tau_{-3}=1$, medium grey area for $\tau_{-3}=10$.
  c) forward shock not radiative: light grey area (we assumed
  $\epsilon_{\rm B}=1$). d) angular smearing time longer than slowing down
  time: right of dotted line.  The hatched area shows how the allowed
  region in parameter space expands if we relax the requirement on $\tau$
  from $\tau_{-3}=1$ to $\tau_{-3}=10$.\label{fig:constraints}}}
\bigskip

It is worth noting that this model makes an exception to the rule that
external shocks cannot produce `Type II' behavior (Sari \& Piran 1997).
This is for two reasons: First, the opening angle of the ejecta is so small
that the angular smearing time is short compared to $\tau$.  Second, the
ambient density is so high that the observed slowing-down time is $\tau
\sim 10^{-3}\,{\rm s}$.  As a result, the total duration of the burst is
determined by the time the central engine operates, while the short term
variability is determined by the mass of the bullets and the statistical
properties of the outflow.  This is an important difference from the
internal shock model, where the variability timescale is set by the
intrinsic time scale of the central engine (e.g., the orbital time in a
merger scenario).

\section{Discussion}
\label{sec:discussion}
In order for the efficiency of the burst to be reasonable, most of the
internal energy produced in the shock must be radiated away immediately
(this requirement holds for {\em all} GRB models).  Electron synchrotron
radiation is the only mechanism remotely efficient enough to produce the
gamma rays. While the efficiency also depends on the transfer of energy
from protons to electrons, we assume here that this process is efficient.
Since we know the observed peak frequency of the gamma rays (of order
$500\,{\rm keV}$, Piran 1999), we can then estimate the radiative
efficiency under the assumption that the gamma rays are produced by
electron synchrotron radiation.  It is usually assumed that the magnetic
field in the shocked gas is in equipartition with the energy density in
relativistic particles.  We therefore parameterize the magnetic field
strength as $U_{\rm B} \equiv {\epsilon_{\rm B}} U_{\rm B, eq}$, where
$U_{\rm B eq}$ is the equipartition magnetic field energy density.

Since the shocked wind material is likely flowing around the bullets at
close to the speed of light (like a cocoon surrounding a radio jet), for
efficient cooling we require that the cooling time scale in the comoving
frame be smaller than the light travel time across the surface of the
bullet $\alpha (R_0 + \Delta R)/\Gamma c$.  If that were not the case, the
material pushed aside by the bullet would cool adiabatically rather than
radiatively.  Independent of $R_0$, this translates to the condition
$\Gamma \leq 1430\cdot [\dot{M}_{-6} {\alpha_{-2}}^2 \epsilon_{\rm
    B}/v_{20}]^{1/4}\sqrt{E_{\nu}/500\,{\rm keV}}$ (to the left of the
light grey area in Fig.\ 2), where $E_{\nu}$ is the observed peak energy of
the gamma rays.  This is not a strict condition, however, since we do not
know what the efficiency of the GRB is.

Since our model assumes a central engine at work (essentially a black box
shooting out bullets at a rate $R(t)$), any distribution of spikes could be
reproduced, since $R(t)$ is arbitrary.  It is, however, surprisingly simple
to reproduce the main features seen in different burst profiles by varying
only a few parameters in our model.  For simplicity, we assume that all the
bullets have the same initial $\Gamma$ and the same sideways expansion
rate, i.e., constant $\alpha$.  We are left with two parameters --- the
number of bullets $N_{100}$ and the average slowing-down time
(eq.\,[\ref{eq:timescale}]) --- and two unknown functions: the mass
distribution of the bullets $N(M_{\rm b})$ and the rate at which they are
released $R(t)$.  We assume that $N(M_{\rm b}) \propto {M_{\rm b}}^{-1/3}$,
chosen to give the observed power spectrum of $P(\tau) \propto
{\tau}^{-1/3}$ (Beloborodov, Stern, \& Svensson 1999).  For $R(t)$ we
assume (for lack of better knowledge) that the bullets are released
randomly over a time interval of 15 sec.  

To produce synthetic GRB lightcurves, we calculated the time dependence of
$\Gamma$ and the associated dissipation rate.  Assuming the bullets are
radiating efficiently and correcting for Doppler boosting and frequency
shifts, we then computed the composite lightcurve for each bullet.  We have
plotted two simulated light curves in Fig.\ 1 for $\tau_{-3} = 10$ and
$N=100$ (panel c) and $N=10^{4}$ (panel d).  It seems that simply by
varying $N_{100}$ and $\tau_{-3}$ we can produce a wide range in light
curve shapes.  More complex features (like the gaps seen in panel a) must
be related to the activity of the central engine and cannot be reproduced
by a random spike rate as assumed above.  We have also plotted the light
curve produced by the deceleration of a single bullet along the line of
sight (Fig.\ 1c, insert).  Note that this profile is very similar to a true
FRED (fast rise, exponential decay) profile.  The rise is instantaneous and
the decay follows a steep power-law (to first order).  While indicative,
these calculations are still rather crude and simplistic.  A more careful
analysis of shotgun GRB lightcurves should be carried out in the near
future.

Afterglows are an important test for any GRB model.  How can we understand
an afterglow in the context of the shotgun model?  It is not immediately
obvious why our model should produce an afterglow at all.  This is because
the bullets are assumed to spread sideways.  As mentioned above, the
bullets will have lost half of their kinetic energy at $\Delta R$, where
they have swept up $1/\Gamma$ of their own mass.  If we simply followed the
dynamics of an individual bullet further in time, it would lose the rest of
its energy exponentially quickly (Rhoads 1997).  This is because the
sideways velocity in the lab frame goes as $1/\Gamma$, so that when
$\Gamma$ starts decreasing, the sideways velocity increases, which in turn
increases the cross-sectional area of the bullet.  As a result, the bullet
can sweep up more mass, which leads to a run-away process.  This would
imply that the ejecta would come to a complete stop not far away from
$\Delta R$ and the observed afterglow would last less than a day.

However, there are many bullets traveling together.  As they expand, they
start increasing the covering factor of the blast.  Once the collection of
bullets reaches unit covering factor, they stop slowing down exponentially,
since further sideways expansion does not lead to an increase in swept up
mass.  As a result, the bullets start traveling collectively, resembling a
collimated blast wave with opening angle $\theta$ rather than a meteor
shower.  The only possible difference between our model and the standard
afterglow models is that in our model, the external density follows a
power-law behavior instead of being constant, which has been discussed by
Dai \& Lu (1998).  Since the opening angle of the merged blast is much
larger than the opening angle of the individual bullets, the sideways
expansion does not affect the dynamics until much later, when the blast has
spread by an angle of order $\theta$ (Rhoads 1997).  This transition from
constant opening angle to rapid sideways expansion has been used in other
models to explain the temporal break seen in the afterglow lightcurve of
GRB990510 (Stanek, et al.\ 1999, Harrison et al.\ 1999, Sari, Piran, \&
Halpern 1999)

It should be noted that we have not attempted to present a specific model
for the production of bullets in this paper.  We are confident that these 
bullets can be produced in GRB outflows and will concentrate on their
production in future work. 

\section{Conclusions}
We have shown that GRBs can be produced by a shower of cold, heavy bullets
shot at bulk Lorentz factor $\Gamma \sim 1000$ into a dense medium.  The
required densities are consistent with a stellar wind from either a blue or
red supergiant.  This ties our model directly to the hypernova scenario.
The gamma rays are produced by the shocks these bullets drive into the
ambient gas.  The total duration of the burst is then determined by the
time the central engine operates rather than the slowing-down time of the
bullets, while the latter produces the short-term variability seen in many
bursts.  After the gamma ray phase (when the bullets have lost half of
their kinetic energy to radiation) the blast waves of the individual
bullets merge into a single collimated shock front, which produces a
standard afterglow in a declining external density profile.  We have made
no attempt to explain how a central engine might produce such a shower of
bullets.  This will be the subject of future work.

\acknowledgements{This research was supported in part by NSF grants
  AST95--29170 and AST98--76887.  MCB also acknowledges support from a
  Guggenheim Fellowship.  We thank Pawan Kumar, Chris Reynolds, Mike Nowak,
  Jim Chiang, Martin Rees, and Annalisa Celotti for helpful discussions.}

\end{document}